\documentclass[conference]{IEEEtran}
\IEEEoverridecommandlockouts
\usepackage{cite}
\usepackage{amsmath,amssymb,amsfonts}
\usepackage{algorithmic}
\usepackage{graphicx}
\usepackage{textcomp}
\usepackage{amsmath,amssymb,amsfonts}
\usepackage{booktabs}
\usepackage[bookmarks=false]{hyperref}
\usepackage[flushleft]{threeparttable} 
\usepackage{xcolor}
\usepackage{enumitem}
\usepackage[skip=3pt]{caption}
\usepackage[framemethod=TikZ]{mdframed}
\usepackage{balance}
\usepackage{xcolor}
\def\BibTeX{{\rm B\kern-.05em{\sc i\kern-.025em b}\kern-.08em
    T\kern-.1667em\lower.7ex\hbox{E}\kern-.125emX}}

\mdfsetup {rightline=false,innerleftmargin=4.5,innerrightmargin=0,innertopmargin=0.5pt,innerbottommargin=0.5pt,
outerlinewidth=0.5pt,topline=false,leftline=true,bottomline=false,
skipabove=.4\baselineskip, skipbelow=2.5pt}

\definecolor{mygreen}{RGB}{0,128,0}

\usepackage{xspace}
\usepackage{xcolor}

\newcommand{\ie}{\textit{i.e.,}\xspace}
\newcommand{\eg}{\textit{e.g.,}\xspace}

\newcommand{\etal}{\textit{et al.}\xspace}

\makeatletter
  \newcommand\tablescript{\@setfontsize\tablescript{8pt}{8}}
\makeatother

\makeatletter
  \newcommand\smalltablescript{\@setfontsize\smalltablescript{7pt}{6}}
\makeatother

\begin{document}

\title{Researcher Bias in Software Engineering Experiments: a Qualitative Investigation}

\author{\IEEEauthorblockN{Simone Romano}
\IEEEauthorblockA{\textit{University of Bari}\\
Bari, Italy \\
simone.romano@uniba.it}
\and
\IEEEauthorblockN{Davide Fucci}
\IEEEauthorblockA{\textit{Blekinge Institute of Technology}\\
Karlskrona, Sweden \\
davide.fucci@bth.se}
\and
\IEEEauthorblockN{Giuseppe Scanniello}
\IEEEauthorblockA{\textit{University of Basilicata} \\
Potenza, Italy \\
giuseppe.scanniello@unibas.it}
\and
\IEEEauthorblockN{Maria Teresa Baldassarre}
\IEEEauthorblockA{\textit{University of Bari}\\
Bari, Italy \\
mariateresa.baldassarre@uniba.it}
\and
\IEEEauthorblockN{\phantom{empty}}
\IEEEauthorblockA{\phantom{empty}  \\
\phantom{empty} \\
\phantom{empty}}
\and
\IEEEauthorblockN{Burak Turhan}
\IEEEauthorblockA{\textit{Monash University and University of Oulu} \\
Melbourne, Australia, and Oulu, Finland\\
burak.turhan@monash.edu}
\and
\IEEEauthorblockN{Natalia Juristo}
\IEEEauthorblockA{\textit{Universidad Politécnica de Madrid} \\
Madrid, Spain \\
natalia@fi.upm.es}
}
\maketitle
\begin{abstract}

Researcher Bias (RB) occurs when researchers influence the results of an empirical study based on their expectations. RB might be due to the use of Questionable Research Practices (QRPs). In research fields like medicine, blinding techniques have been applied to counteract RB. We conducted an explorative qualitative survey to investigate RB in Software Engineering (SE) experiments, with respect to: \textit{(i)}~QRPs potentially leading to RB, \textit{(ii)}~causes behind RB, and \textit{(iii)}~possible actions to counteract RB including blinding techniques. Data collection was based on semi-structured interviews. We interviewed nine active experts in the empirical SE community. We then analyzed the transcripts of these interviews through thematic analysis. 
We found that some QRPs are acceptable in certain cases. Also, it appears that the presence of RB is perceived in SE and, to counteract RB, a number of solutions have been highlighted: some are intended for SE researchers and others for the boards of SE research outlets. 

\end{abstract}


\section{Introduction}
In research, \textit{bias} is defined as the combination of various design, data, analysis, and presentation factors tending to produce findings that should not be produced~\cite{Ioannidis:2005}. \textit{Researcher Bias} (\textit{RB}), or \textit{experimenter bias}, occurs when researchers, consciously or unconsciously, influence the results of an empirical study based on their expectations.
In some cases, RB is due to the adoption of Questionable Research Practices (QRPs) to follow one's agenda and achieve specific expectations---\eg adjusting an experimental sample until statistically significant results are found. 
Another form of bias is \textit{publication bias}, which occurs when studies are published based on their results---usually positive results are more likely to be published than negative ones~\cite{DAA08}.
To counteract RB, according to established guidelines in Software Engineering (SE), researchers should disclaim their stance regarding an outcome. For example, Wohlin \etal~\cite{Wohlin:2012} consider \textit{experimenter expectancies} as a social threat to construct validity.
Nevertheless, it has been shown that RB affects SE studies~\cite{Shepperd:14,Jorgensen:2016}.

In this paper, we report the opinions of a group of experts about themes related to RB in SE experiments. To this end, we conducted a qualitative explorative survey. The scope of our research is both on human- and technology- oriented experiments~\cite{Wohlin:2012}. We are concerned with: QRPs related to RB, causes behind RB, and possible actions to counteract it. Regarding the latter, we focus on two techniques---\textit{blind data extraction} and \textit{blind data analysis}.
The former consists of hiding some information (\eg treatment assignment) from the researchers who extract the data; whereas, the latter is the temporary and judicious removal of labels and alteration of values before someone analyzes the data~\cite{MP15}.
Although extensively used in other research fields like medicine and physics~\cite{Karanicolas:2010,MP15}, SE researchers have used these techniques in few occasions~\cite{Fucci:16,Sigweni:2015}.
To collect data, we conducted semi-structured interviews. Nine experts in the field of empirical SE took part in these interviews. We then applied thematic analysis~\cite{King:2004} to organize experts' opinions. 


\section{Background}\label{sec:background}
In this section, we survey the status of RB in SE, along with QRPs associated to it. Moreover, we describe countermeasures to RB including blinding techniques, which our survey explicitly focuses on.

\subsection{Questionable Research Practices and Researcher Bias}\label{sec:QRPs}
Cases of QRPs, exploiting the gray area of what is considered acceptable, have been mounting in medicine, natural sciences, and psychology (\eg~\cite{FCI17,JLP12}).
As for the SE research field, J{\o}rgensen \etal~\cite{Jorgensen:2016} documented the presence of RB and publication bias in SE experiments. The authors conducted a quantitative questionnaire-based 
survey, with researchers from some SE sub-communities, comprising questions about QRPs potentially leading to RB and publication bias. Three out of seven questions were on QRPs related to RB, namely: 
\begin{enumerate}[leftmargin=*]
    \item \textit{Post-hoc hypotheses}---defined as reporting the results of one (or more) hypothesis tests where at least one of the hypotheses is formulated after looking at the data.  
    \item \textit{Post-hoc outlier criteria}---defined as developing or changing the rules for excluding data (\eg outlier removal) after looking at the impact of
doing so on the results.
    \item \textit{Flexible reporting of measures and analysis}---defined as using several variants of a measure or several tests and then reporting only the measures and tests that give the strongest results.
\end{enumerate}
The results suggest that: \textit{(i)} 67\% of the  respondents had followed the post-hoc hypotheses practice; \textit{(ii)} 55\% had followed the post-hoc outlier criteria practice; and \textit{(iii)} 69\% had followed the flexible reporting of measures and analysis practices. J{\o}rgensen \etal~\cite{Jorgensen:2016} also built a model---150 randomly-sampled SE experiments fed the model---to estimate the proportion of correct results at different levels of RB and publication bias. The model suggests that both RB and publication bias affect SE experiments since 52\% of the statistically significant tests do not match a situation with no or low RB and publication~bias. Shepperd \etal~\cite{Shepperd:14} in their meta-analysis of defect prediction techniques came to a conclusion similar to that by J{\o}rgensen \etal~\cite{Jorgensen:2016}. The authors pointed out the presence of RB in the studies included in the meta-analysis as the factor with the largest effect was the research group publishing the paper, while the effect of the prediction technique was small.

\subsection{Countermeasures to Researchers Bias}\label{sec:countermeasure}
Potential solutions to counteract RB have been proposed~(\eg~\cite{Nuz15,PWD10}). We can group them into: \textit{(i)}~\textit{rival theories}; \textit{(ii)}~\textit{transparency}; and \textit{(iii)}~\textit{blinding}.
The first category consists of considering alternative or competing hypotheses with respect to the ones being tested in the study. 
The researcher should then devise experiments that can explicitly distinguish competing hypotheses and, if possible, develop experiments that can distinguish between alternative theories. The researcher should collaborate with a \textit{team of rivals}---\ie other researchers that, while being skeptical about the hypotheses, collaborate towards developing alternative explanations. 

Several approaches fall under the umbrella of the transparency category.
The prime example is \textit{open science}, \ie  
 the practice of sharing research data, computer code, and lab packages for public scrutiny so attempting to reproduce results.
In fields like medicine or psychology, transparency is also achieved through \textit{pre-registration} (or \textit{registered report}).
It consists of submitting a paper presenting the study rationale and planning for peer review before its conduction.
Once the paper is accepted, the researchers can conduct the study and submit a paper with the obtained results for a second round of revision. The paper cannot be rejected due to the study results (\eg negative results) while it can be for other reasons (\eg deviations from the pre-registered analysis procedure)~\cite{Nosek:2014}.

Finally, blinding (or \textit{masking}) means concealing research design elements (\eg treatment assignment or research hypotheses) from individuals involved in an empirical study (\eg participants, data collectors, or data analysts)~\cite{Miller:2011,Page:2013}.
The use of blinding techniques has been encouraged fields like medicine and physics~\cite{Karanicolas:2010,MP15}. As for SE, Shepperd \etal~\cite{Shepperd:14} have fostered researchers to use blinding techniques in their studies; however, few researchers have applied blinding in SE studies so far, namely: Fucci \etal~\cite{Fucci:16}, who used blind data extraction and analysis in a human-oriented experiment; and Sigweni and Shepperd~\cite{Sigweni:2015}, who applied blind data analysis in a technology-oriented experiment. 

To explain how blind data extraction and analysis work, let us take as an example the experiment by Fucci \etal~\cite{Fucci:16}. The study goal was to investigate the claimed effects of Test-Driven Development (TDD), such as the increase in developers' productivity.
The experiment compared a \textit{treatment group}---\ie developers who applied TDD to implement some programs---to a \textit{control group}---\ie developers who implemented the same programs as the other group but by applying Test-Last Development (TLD).
After the experiment took place, the raw dataset (\ie the programs the developers implemented) was given to a researcher that played the role of data extractor.
Given the raw dataset, this researcher extracted the values of the metrics (\eg the PROD metric quantified developers' productivity) so obtaining the dataset.
The extraction of the metrics was done blindly because the data extractor was not aware of the experimental goal, hypotheses, treatment assignment, and design.
The dataset was then given to the data analysts who performed the blind data analysis.
They were not aware of the experimental goal and they worked on a sanitized dataset---\ie
the values of the independent variables were temporarily replaced (\eg the TDD and TLD groups became the A and B groups, respectively) and the dependent variables 
were temporarily anonymized (\eg PROD was renamed as DV1).
To correctly analyze the data, the analysts were provided with a minimal description of the dependent and independent variables (\eg DV1 assumed values in $[0,1]$)
as well as the experimental design in which some information was adequately hidden (\eg the experimental groups were referred as A and B).
One the analysis was complete, the hidden information was revealed. 

\section{Study Design and Limitations}\label{sec:interviews}

\subsection{Goal}
Unlike quantitative research that seeks to provide quantifiable responses to some research questions, qualitative research (like ours) is concerned with understanding subjects' viewpoints about a given phenomenon and discovering the causes behind that phenomenon as noticed by the subjects~\cite{Wohlin:2012}. Therefore, the goal of our study is to elicit the opinions of a group of experts about RB in SE experiments, including: \textit{(i)}~QRPs potentially leading to RB; \textit{(ii)}~causes behind RB; and \textit{(iii)}~potential actions to counteract RB with a focus on, but not limited to, blind data extraction and analysis. That is to say that, while the studies by Shepperd \etal~\cite{Shepperd:14} and J{\o}rgensen \etal~\cite{Jorgensen:2016} show the presence of RB in SE studies by leveraging quantifiable responses, we are more interested in investigating the phenomenon of RB based on researchers' viewpoints. Our study can be thus considered complementary to those above-mentioned~\cite{Shepperd:14,Jorgensen:2016}.

\begin{table*}[t]
\centering
\caption{Characterization of the interviewees involved in this study.}
\label{tab:interviewees} 
\smalltablescript
\begin{tabular}{@{}llllll@{}}\toprule
ID & Institution region & Academic position & Main research interest & Experience as experimenters & Last published experiment\\
\midrule
R1 & Southeastern Europe & Assistant professor & Defect prediction & 5-10 (years) & $<$ 6 months \\ 
R2 & Northern Europe & PhD student         & Human and social aspects of SE & 1-5 (years) & $<$ 18 months \\ 
R3 & Northern Europe & Full professor      & Mining software repositories & 11-20 (years)   & $<$ 6 months \\ 
R4 & Northern America & Associate professor  & Agile software development & 11-20  (years) & $<$ 6 months\\ 
R5 & Central Europe & Assistant professor & Software maintenance and evolution & 5-10 (years) & $<$ 3 years \\ 
R6 & Southern Europe & Associate professor & Software economics and metrics & 11-20 (years) & $<$ 1 year \\ 
R7 & Southern Europe & Assistant professor          & Project and process management & 11-20 (years)    & $<$ 1 year \\ 
R8 & Southern Europe & Full professor      & Collaborative software development  & $>$ 20  (years) & $<$ 18 months \\ 
R9 & Southern Europe & Full professor      & Software economics and metrics & 11-20  (years)    & $<$ 6 months \\ 
\bottomrule
\end{tabular}
\vspace{-2em}
\end{table*}

\subsection{Protocol}
We planned a series of interviews with experts from the empirical SE community to investigate on RB.
Despite interviewees are time-consuming, we opted for this data collection means, rather than questionnaires, because: \textit{(i)}~it allows achieving higher response rates; \textit{(ii)}~it decreases the number of ``don't know'' and ``no answers''; and \textit{(iii)}~the interviewer can ask for clarifications if needed~\cite{Wohlin:2012}. Also, such a method fits the explorative intention of our study. 

We recruited researchers in our network based on their experience in SE experimentation (both human- and technology- oriented).
Nine researchers were available for an interview either face-to-face or by phone. 
Each interview involved the same interviewer (\ie DF) together with one interviewee at a time. 
We obtained consent from the interviewee to be audio-recorded and we informed each of them that the gathered data would be treated confidentially. 
Each interview lasted between 50 and 75 minutes. 
We used semi-structured interviews~\cite{Wohlin:2012}. That is, the questions listed in the interview script were not necessarily asked in order because, depending on how the conversation evolved, some questions were handled before than others. Semi-structured interviews enable improvisation and exploration of the investigated phenomenon. 
The interview script serves to guide the discussion and make sure that relevant topics are covered~\cite{Wohlin:2012}.

The interview script\footnote{\url{https://doi.org/10.6084/m9.figshare.12356213.v1}} consisted of eight parts. 
The objective of the first part (\ie \textit{Warm Up}) was to gather demographic information on the interviewees (\eg where the researcher was employed or her research interests). 
This information allowed us to characterize the study context. As for the second part (\ie \textit{Experiments}), the interviewer asked to guide him through the usual experimental process of the interviewee. The goal was to break the ice between interviewer and researcher by gathering information on how researchers design an experiment and the division of work in case a team of researchers is involved.
We were also interested in their perception of threats to validity that can arise given their design choices. In the third part (\ie \textit{QRP}), we  gathered the interviewees' viewpoints on some QRPs recently reported in the survey by J{\o}rgensen \etal~\cite{Jorgensen:2016}.
We considered those QRPs potentially leading to RB, namely: post-hoc hypotheses, post-hoc outlier criteria, and flexible reporting of measures and analyses.
We did not consider QRPs related to publication bias because our paper does not focus on this kind of bias. In the fourth part (\ie \textit{RB}), we focused on how the interviewees perceived RB in SE experiments, the causes behind it, and their suggestions to avoid/mitigate it. With the fifth (\ie \textit{Blind Data Extraction}) and sixth (\ie \textit{Blind Data Analysis}) parts, we centered the discussion on blind data extraction and analysis, respectively.
We gathered  opinions about the aforementioned techniques to cope with RB. As for the seventh part (\ie \textit{Blind Data Extraction and Analysis}), we focused on the interviewees' thoughts on the use of blind data extraction and analysis together, as well as how to foster the use of these techniques. We ended each interview (\ie \textit{Wrap Up}) by asking whether the interviewee would use blind data extraction and analysis in her future experiments.

\subsection{Participants}
All the interviewees had, at the time of our study, published at least one experiment in one of the SE higher quality venues (\ie ICSE, EMSE, IEEE TSE, and ACM TOSEM).
In Table~\ref{tab:interviewees}, we report some interviewees' information gathered in the \textit{Warm Up} part of the study. We guarantee the anonymity of the interviewees by referring to each of them through an ID (from R1 to R9). As Table~\ref{tab:interviewees} shows, the participates were quite heterogeneous in terms of location of their institution, academic position, main research interest, years of experience as experimenter, and date of last published experiment.

\subsection{Data Analysis}
After transcribing the recordings of the interviews, we (\ie SR, MTB, and GS) analyzed the transcripts using a thematic analysis approach called template analysis, which is known to be flexible and fast~\cite{King:2004}. Template analysis allows developing a list of codes, each of which identifies a theme within the transcripts. The codes are arranged in a \textit{template}---it usually is a hierarchical structure of codes---showing the relationships among themes, as defined by the investigators. In template analysis, the investigators start analyzing the transcripts by using an initial template. That is, they  attach pre-defined codes, arranged in a template, to delimit portions of text related to themes. As King~\cite{King:2004} suggests, the best starting point for developing an initial template is the interview script. Accordingly, we developed our initial hierarchical template (see the non-bold text in Figure~\ref{fig:template}) from the interview script.
As customary in template analysis, we revised the initial template during the analysis~\cite{King:2004}. In particular, we renamed the second-level code \textit{Presence of Researcher Bias} as \textit{Presence of Researcher Bias and Clues} because we found portions of text about clues suggesting the presence of RB.
We concluded the analysis when any portion of text relevant to the goal of our study was coded and we agreed on the obtained template. To ease the thematic analysis, 
we used the \textit{ATLAS.ti} tool.

\begin{figure}[t]
\center
\smalltablescript
\begin{tabular}{@{}p{9px}p{4px}p{1px}p{180px}@{}}\toprule
\multicolumn{4}{@{}l}{Experiments}  \\
                & \multicolumn{3}{@{}l}{Researcher Roles} \\
                & \multicolumn{3}{@{}l}{Threats to Validity}\\
\multicolumn{4}{@{}l}{Questionable Research Practices} \\
                & \multicolumn{3}{@{}l}{Post-hoc Hypotheses}\\
                & \multicolumn{3}{@{}l}{Post-Hoc Outlier Criteria}\\
                & \multicolumn{3}{@{}l}{Flexible Reporting of Measures and Analyses}\\
\multicolumn{4}{@{}l}{Researcher Bias}\\
                & \multicolumn{3}{@{}l}{Presence of Researcher Bias \textbf{And Clues}}\\
                & \multicolumn{3}{@{}l}{Causes of Researcher Bias}\\
                & \multicolumn{3}{@{}l}{Coping with Researcher Bias}\\
                &   &   \multicolumn{2}{@{}l}{Blind Data Extraction} \\
                &   &   & Usefulness of Blind Data Extraction \\
                &   &   & Drawbacks of Blind Data Extraction \\
                &   &   \multicolumn{2}{@{}l}{Blind Data Analysis} \\
                &   &   & Usefulness of Blind Data Analysis \\
                &   &   & Drawbacks of Blind Data Analysis  \\
                &   &   \multicolumn{2}{@{}l}{Blind Data Extraction and Analysis} \\
                &   &   & Effectiveness of Blind Data Extraction and Analysis \\
                &   &   & Fostering Blind Data Extraction and Analysis        \\ \bottomrule
\end{tabular}
\caption{Initial and final templates (in bold, text added to the initial template to obtain the final one).}\label{fig:template}
\vspace{-2em}
\end{figure}

\subsection{Limitations}
When interpreting the findings from qualitative investigations, some limitations have to be taken into account: \begin{itemize}[leftmargin=*]
    \item The interviewees may not answer truthfully because, for example, they are scarcely motivated or afraid of being judged. To mitigate this threat, the participation in the study was voluntary---volunteers are generally more motivated~\cite{Wohlin:2012}---and we informed the interviewees about data confidentiality.     
    \item The number of the interviews might threaten the validity of results. However, Guest \etal~\cite{Guest:2006} observed that a sample of six interviews may be sufficient to allow development of meaningful themes and useful interpretations. Given the results observed after the analysis of the nine interviews, we believe to have hit a point of diminishing return~\cite{Guest:2006} for which increasing the number of interviews will unlikely generate more evidence. Moreover, our plan includes (quantitative) surveys, based on questionnaires, with researchers from different SE sub-communities (\eg ICSE or ESEM) to understand how much they agree with the interviewees' statements (\ie we are going to apply \textit{methodological triangulation}~\cite{Thurmond:2001}) and whether or not there are differences 
    among different SE sub-communities.
    \item Our findings might not generalize to researches sampled from a different population. As previously mentioned, we are going to investigate this point in our long-term plan with quantitative surveys.
    \item The investigator might unconsciously influence the results based on its expectations. We mitigated such a threat by involving more people in conducting and analyzing the interviews. In particular, DF was the interviewer while SR, MTB, and GS performed the data analysis (\ie we applied \textit{investigation triangulation}~\cite{Thurmond:2001}).
\end{itemize}


\section{Findings}\label{sec:results}
We present the findings emerging from the interviews based on the main themes identified by the first-level codes (\eg \textit{Questionable  Research  Practices}) of the final template shown in Figure~\ref{fig:template}. To bring credibility to our findings, we present them together with some excerpts from the transcripts.

\subsection{Experiments}
As shown in Figure~\ref{fig:template}, two sub-themes were defined within this main theme: the roles of researchers in SE experiments and how they cope with threats to validity in their experiments.   

\textbf{Researcher Roles.} It emerged that, when conducting an experiment, there is a division of roles among the researchers. 
Each researcher covers one or more roles (\eg one researcher is involved in the planning of the experiment and in its execution, another one extracts the metrics from the raw data and so on). 
However, it seems that only one researcher takes care of data analysis (\ie one researcher plays the data analyst role). An excerpt from the interview with R6 follows:  

\begin{mdframed} 
\tablescript{We [our research group] outlined the experiment design. The researchers from [other country] translated the experiment material into [other language] and carried out the experiment in [other country]. We then received the gathered data, some Excel files, and one of us executed~the~analysis.}
\end{mdframed}


\textbf{Threats to Validity.}
When we asked the interviewees to elaborate on threats to validity, they provided a number of examples, but none mentioned RB. 
\subsection{Questionable Research Practices}
This  theme includes three sub-themes (see Figure~\ref{fig:template}): the participants' perceptions of post-hoc hypotheses, post-hoc outlier criteria, and flexible reporting of measures and analyses.

\textbf{Post-hoc Hypotheses.} According to the interviewees, post-hoc hypotheses should not to lead to RB as long as: \textit{(i)} the researchers clearly report that such hypotheses are formulated in retrospect; or \textit{(ii)} it is possible to ground such hypotheses on prior work (thus, there is no need to make clear that such hypotheses are post-hoc). Regarding \textit{(i)}, R5~said: 
\begin{mdframed}
\tablescript{In this case, first of all I am not sure we can talk about formulating hypotheses because you are already looking at the data of an experiment [...] In general, I don't think there is anything wrong with that if, and I think it is completely sound, if you explicitly say that it is an unexpected result when reporting this result. This is different from saying ``we wanted to investigate this and we found that it is supported by the~data.''} 
\end{mdframed}
As for the point \textit{(ii)}, R3 said:
\begin{mdframed}
\tablescript{Of course, there's the fact that, the hypothesis should be grounded on prior work. If you can ground something to solid prior work, then it doesn't really matter whether it was sort of after the fact.}
\end{mdframed}
Also, it seems post-hoc hypotheses could be a means to get new insight into the studied phenomenon, which researchers had not thought about when the study was planned.~R4~said:
\begin{mdframed}
\tablescript{It [a post-hoc hypothesis] emerged from the data and inevitably happens. When you look at the data, you may have, you may think of new insights that you haven't thought about because there is information that was not anticipated.
[...] Sometimes there are research methodologies that don't even assume any questions, they are completely totally exploratory. So let's suppose that you have a set of questions, and you wanna answer them first. After you answer those questions, then you see some other patterns in your data and then, in the next iteration, you formulate a set of other questions that maybe you can answer based on the same data. This is completely okay but it's not the same as~fishing. }
\end{mdframed}

\textbf{Post-hoc Outlier Criteria.}
The interviewees seem to believe that this practice should be avoided because it potentially leads to RB, though not necessarily. R5 told us: 
\begin{mdframed}
\tablescript{Looking at the results and then removing  outliers could sometimes be sensible, but I think the bias would be too~strong. }
\end{mdframed}
In case researchers apply the post-hoc outlier criteria practice, the interviewees agreed that they should declare the use of this practice in the paper by providing, for example, the following information: \textit{(i)} the results before and after removing outliers; \textit{(ii)} the reasons behind the outlier removal; and \textit{(iii)} an interpretation of the results (\eg why, after the outlier removal, a null hypothesis passes from non-rejected to rejected). R4~said:  
\begin{mdframed}
\tablescript{As long as you declare the results and you present maybe both of them [before and after the outlier removal], depending on how other factors influence your interpretation. Maybe there are other things that you discovered during your data analysis that justifies that decision. But as long as you declare them, I mean that is one of the purposes of the peer review, the reviewers can also decide which one is, whether that decision was sensible or~not.}
\end{mdframed}


\textbf{Flexible Reporting of Measures and Analysis.}
Based on interviewees' experience, when researches can choose among equivalent statistical hypothesis tests (\eg t-test or F-test), the results (\ie p-values) are not so different. R8 told us:
\begin{mdframed}
\tablescript{It's true that there are a lot of statistical hypothesis tests and there are a lot of variants as well, when using statistical packages we are spoilt for choice, but in my experience they don't vary so~much.
}\end{mdframed}
Furthermore, according to R3, if a statistical hypothesis test revealed a significant difference 
that an equivalent test did
that difference would be probably negligible. In other words, the effect size would show the true impact of that difference, so having or not a significant difference would not matter:   
\begin{mdframed}
\tablescript{It [using a statistical hypothesis test or an equivalent one] doesn't really impact the results very much. It's a very very tiny difference, at least what I have seen. It doesn't change from .04 to .0004, or something. I mean you might, if you again use this magical threshold of .05, then it might matter. But if you report the effect sizes, then it really doesn't. The effect sizes sort of reveal the true~impact. }
\end{mdframed}
As for the practice of using several variants of a measure and then reporting only the variants that give the strongest results, it is perceived as a bad practice. The researchers should discuss any variant of that measure in the paper. R4 said:
\begin{mdframed}
\tablescript{Yeah I think that is a no, in general. If you've done [flexible reporting of measures], there needs to be a discussion of how your attempt to triangulate the results with different measures failed. That should be part of the discussion and it's part of the validity threats that you have. 
}    
\end{mdframed}

\subsection{Researcher Bias}
This theme has three sub-themes (see Figure~\ref{fig:template}): the presence of RB in experiments and clues suggesting such a presence; causes of RB; and strategies to cope with RB.

\textbf{Presence of Researcher Bias and Clues.}
From the interviews, it emerged that RB affects the SE community. Although the interviewees did not have proofs about the presence of RB in SE, they pointed out four clues suggesting its presence: \textit{(i)}~RB affects any community (\eg medicine or psychology); \textit{(ii)}~when reviewing papers, it is not rare to suspect authors biasing the results; \textit{(iii)}~whoever could unconsciously bias the results based on her expectations; and \textit{(iv)}~there are sometimes inconsistent results among studies investigating the same constructs. On the points \textit{(i)} and \textit{(ii)}, R4~stated: 
\begin{mdframed}
\tablescript{I think it [RB] must be happening because it's probably happening in every community. But I'm not sure. I mean I think, in terms of my review work, when things are suspicious, it's usually obvious and it's usually not just from one reviewer picking on them, rather, multiple reviewers do and it's only because, the researchers actually let it be understood in the~paper.
}\end{mdframed}
As for the point \textit{(iii)}, R3's thought follows:
\begin{mdframed}
\tablescript{I guess everyone that does experiments is somehow biased because you know that negative results cannot be published and it probably, sort of unconsciously, alters your actions.
}\end{mdframed}
On the last point, R8 said:
\begin{mdframed}
\tablescript{That is, if I see that a given result isn't confirmed [by another study], then it is a clue of researcher~bias.} 
\end{mdframed}

\textbf{Causes of Researcher Bias.}
Four causes of RB emerged from the interviews. First, interviewees believed that \textit{negative-results papers are usually rejected}. This would lead researchers to bias their results (\eg transforming non-significant results into statistically significant ones). R2 said: 
\begin{mdframed}
\tablescript{I think the main reason to that is there is no acceptance for reporting the negative results. You are a researcher and your responsibility is just to explore the phenomenon, whether it is in favor of your hypothesis or it's against your hypothesis you should report it, but I've personally felt like there is no in general acceptance for that.
}\end{mdframed}
Second, the \textit{pressure of publishing papers} can lead researchers to (unconsciously or consciously) bias the results. R5~said:     
\begin{mdframed}
\tablescript{Especially young researchers, for example Ph.D. students, that carry out and are therefore responsible for the experiment, may tend to have high expectations on what they have developed or towards the hypothesis being verified, to the point that, even unconsciously, they may tend to guide the experiment towards a certain expected result. I am quite confident to say that, although not always, this occurs especially with novice experimenters that are more eager for publications and may therefore be led to experimenter~bias. 
}\end{mdframed}
Third, it seems that \textit{revision processes of SE conferences/journals are focusing too much on the empirical assessment}, rather than on the contributions of the ideas to the body of knowledge. Thus, researchers would be led to bias their studies by making the results more publishable. R5 told~us:
\begin{mdframed}
\tablescript{I think that the main problem of several review processes is that they are highly based on the empirical aspect and much less on the novelty of the ideas. So in spite of you propose an interesting and novel idea that several other researchers can build on, if the experimental results are not strong enough you are likely to receive a comment like ``okay nice idea but ...''. On the other hand, if a study is empirically perfect, from the point of view of the design and results, but has very limited novelty, it's difficult that it will be rejected.
}\end{mdframed}
Fourth, the \textit{immaturity of the SE field and its researchers}. That is, some researchers believe not to bias the results of their experiments when they actually do. In this respect, R9~said:
\begin{mdframed}
\tablescript{Sometimes, in good faith, one may think that this does not represent an actual threat to the experiment. 
}\end{mdframed}

\textbf{Coping with Researcher Bias.} 
The interviewees suggested seven strategies to cope with RB. First, the use of \textit{pre-registration} in SE conferences/journals (see Section~\ref{sec:countermeasure}). This should prevent negative-results papers from being rejected. Also, pre-registration increases both credibility of study results and study replicability~\cite{Nosek:2014}. Therefore, researchers should be less prone to bias their results. On this point, R5 said:
\begin{mdframed}
\tablescript{Personally, I have an idea. It doesn't relate to the experimental design, rather to a discipline. It consists of having dedicated tracks of a conference or sections of a journal where authors don't submit the results of an experiment, but the experiment they plan to carry~out. 
}\end{mdframed}
Second, fostering \textit{open data policies} in SE conferences/journals. This means not only making the gathered data publicly available, but also the analysis scripts of the study. Such open data policies should allow the reviewers (and any other researcher) to repeat the data analysis of that study so attributing credibility to study outcomes and increasing the replicability of the study. Therefore, researchers should be discouraged from biasing their studies. 
R1's thought follows:  
\begin{mdframed}
\tablescript{Another thing could be publishing all the analyses together with the data. But then that implies during the review process that, as a reviewer, I have to go and take a look at the analysis as~well.
}\end{mdframed}
Third, \textit{duplicate data analysis}. That is, two researchers analyze the same data with their own scripts without interacting with one another. Then they exchange the scripts and data to cross-check them. Finally, the results of the data analysis are compared. R5 mentioned this kind of data analysis, (she/he was using at the time of the interview), which should mitigate the unconscious bias of researchers involved in the data analysis.   
\begin{mdframed}
\tablescript{The only thing I do, from about three years, is that data is always analyzed independently by two researchers. Next, they exchange the scripts and cross-check them. They exchange the data and cross-check them as well. Finally, they compare their~conclusions.  
}\end{mdframed}
Fourth, \textit{means for increasing the awareness} of RB in SE. For example, panels on RB in SE, an ethical code for SE warning researchers against this kind of bias, or papers on RB in SE. Therefore, by increasing the awareness of RB, researchers should be warned against this kind of bias. 
R6~told: 
\begin{mdframed}
\tablescript{Fostering panels and discussions on this [researcher bias], conducting surveys and studies, like the one you are conducting, to understand the status of the community.}
\end{mdframed}
Fifth, \textit{guidelines for reviewers} in SE conferences/journals. These guidelines should instruct the reviewers not to judge papers on the basis of the study results (\ie positive/negative results). As a consequence, researchers would bias the study results less because having a paper reporting positive/negative results would be equally valid.   
On this matter, R4 said:
\begin{mdframed}
\tablescript{Perhaps review guidelines may also help, in the sense that you instruct the reviewers, specifically not to bias their reviews only if the results are favorable to the hypothesis of the researchers.
}\end{mdframed}
Sixth, \textit{ad-hoc research tracks} in SE conferences (or ad-hoc issues in SE journals). For example, specific tracks for papers reporting negative results or specific tracks for studies having a not as strong empirical assessment. Such kind of tracks should lead researchers not to bias their results to have more publishable results. 
On this point, R7 said: 
\begin{mdframed}
\tablescript{Having various publication-levels where non-rigorous studies carried out by research groups or companies can be published in prestigious journals. 
}\end{mdframed}
Seventh, \textit{replicated experiments} because the more the results of a study are confirmed by replications, the lower the likelihood of RB is. 
In this respect, R8 told us: 
\begin{mdframed}
\tablescript{I trust when the results are confirmed by more studies carried out by researchers that are not co-authors. I don't think only one paper is enough. I don't confide in the results of only one paper. Of course, this doesn't mean that single studies are conducted incorrectly or are error-prone, it simply impacts on generalizability.} 
\end{mdframed}
Besides the strategies, mentioned by the interviewees, to cope with researches bias, we asked their thoughts on two further strategies, \ie blind data extraction and blind data analysis, used alone or together. In the following subsections, we report the findings concerning the sub-themes for blind data extraction, blind data analysis, and both these strategies.  

\subsubsection{Blind Data Extraction}
Two sub-themes were defined for this theme (see Figure~\ref{fig:template}): usefulness and drawbacks of blind data extraction in SE experiments.  

\textbf{Usefulness of Blind Data Extraction in SE.} It emerged from the interviews that blind data extraction could be a useful technique to mitigate RB because, even when extracting the metrics, a researcher could favor a given treatment  based on her expectations. In other words, if the data extractor (\ie the person who is responsible of extracting the metrics from the raw dataset) is aware of research design elements (\eg treatment assignment), then the likelihood of influencing the results towards a given treatment is higher. This is why having blinded extractors would lessen the likelihood of influencing the results.
In this respect, R3~said: 
\begin{mdframed}
\tablescript{Yeah, I think it [blind data extraction] sounds like a good idea. I believe that they [the researchers] may apply bad practices of statistical analysis but actually I believe more that one does it, consciously or unconsciously, while they code the data, or do it even before running the experiments because the researcher knows what treatment is and what the control is. I think that's a good idea that labels are removed and someone else transforms the data. }
\end{mdframed}

\textbf{Drawbacks of Blind Data Extraction.} As for the drawbacks of blind data extraction, the interviewees pointed out that the implementation of blind data extraction requires at least two people: an individual (\ie the study executor) responsible of executing the experiment and another individual (\ie the data extractor) with the necessary skills to extract the metrics from the raw dataset. The latter has to be blinded to research design elements. This seems to be less feasible when both study executor and data extractor belong to the same research group---guessing or finding out about hidden information (\eg research hypotheses) would be more likely when both executor and extractor belong to the same research group. Therefore, to implement blind data extraction, it is preferable to have: \textit{(i)}~a research collaboration between two research groups where the experimenter and the extractor are not part of the same group, or \textit{(ii)}~an external expert 
that takes care of the metric extraction. To this respect, R8 stated:
\begin{mdframed}
\tablescript{I think it [blind data extraction]'s complicated. In many cases it's you and your PhD student, do you really think that your student isn't aware of who did certain things? [...] Maybe it can work in a joint experiment where you have a large group of people collaborating from various independent research groups. On the other hand, within the same group it is applicable in theory because you have several researchers involved, however it becomes an ``open secret'' as everyone is aware of what is going on. How much would it work within the same group?}
\end{mdframed}
Note that R5 had already used blind data extraction. She involved some experts to extract metrics from a raw~dataset:
\begin{mdframed}
\tablescript{Well now that you have mentioned it [blind data extraction], we actually have done it on two papers in the past that I had forgotten about. What we did was to gather the artifacts produced by the participants and then give all to external people who evaluated the artifacts. [...] Yes, I think this is surely~useful. 
}\end{mdframed}

\subsubsection{Blind Data Analysis}
Two sub-themes were defined for this theme (see Figure~\ref{fig:template}): usefulness and drawbacks of blind data analysis in SE experiments. 

\textbf{Usefulness of Blind Data Analysis.} 
Blind data analysis seems a useful technique to mitigate RB. A blinded analyst (\ie an analyst unaware of research design elements) would perform the data analysis more objectively than an analyst aware of research design elements. In this respect, R7 said: 
\begin{mdframed}
\tablescript{It can be a means for a more objective analysis because it's human to be inclined to one's proposals and expectations. This can be thus an involuntary contribution, either positive or negative, that a researcher~provides. 
}\end{mdframed}

\textbf{Drawbacks of Blind Data Analysis.}
The drawback of blind data analysis is that at least two researchers are needed---the former one conducts the study and sanitized the dataset, while the latter one performs the data analysis on the sanitized dataset. Moreover, it is preferable 
that the researchers do not belong to the same research groups. For example, R8~said:
\begin{mdframed}
\tablescript{It's similar to blind data extraction. That is, if you are conducting a joint experiment, you can apply blind data analysis.}
\end{mdframed}

\subsubsection{Blind Data Extraction and Analysis} We defined three sub-themes for this theme: effectiveness of blind data analysis and extraction in coping with RB, strategies to foster the adoption of blind data analysis and extraction in SE experiments, and intention to use blind data analysis and extraction. 

\textbf{Effectiveness of Blind Data Extraction and Analysis.} 
RB could arise even if blind data extraction and analysis are applied together. That is, using both blind data analysis and extraction is considered a way to mitigate RB. 
In fact, RB could arise not only during the metric extraction and analysis phases but also during the execution of the experiment itself. In this respect, we report R3's answer when we asked if the combination of blind data extraction and blind data analysis was enough to cope with RB:    
\begin{mdframed}
\tablescript{Most likely not. Like I said previously, the step before where you set up and where you run the experiment also introduces some~[bias].}
\end{mdframed}

\textbf{Fostering Blind Data Extraction and Analysis.} 
The interviewees suggested a number of strategies to ease the adoption of blind data extraction and analysis in SE. The first strategy is a \textit{policy} for conferences/journals similar to the double-blind peer-review one. That is, this policy would consist of requiring that any submitted experiment to that conference/journal had to use blind data extraction and analysis. However, this strategy is not always feasible, as the same interviewees observed, due to the following reasons: \textit{(i)}~the reviewers cannot make sure the authors of a paper have really used blind data extraction and analysis; \textit{(ii)}~researchers, who are not involved in research collaborations, would be harmed by this policy; and \textit{(iii)}~empirical evidence on the effectiveness of blind data extraction and analysis in SE is necessary to foster conferences/journals to adopt this policy. 
Regarding point \textit{(ii)}, R8 said:  
\begin{mdframed}
\tablescript{In most cases, you have a [research] group that works independently... it does not involve several units, or you have a group made up of Ph.D. student and supervisor. In this case, how do you distinguish the roles and introduce any blinding in the process? } 
\end{mdframed}
As for the last point, R4 said:
\begin{mdframed}
\tablescript{The conference committees won't do it [that policy] without any evidence that it's gonna be effective, just because it sounds like a good idea. Then, if there is enough evidence that it's a good idea, then maybe some conferences will start using it [that policy].}
\end{mdframed}
The second strategy to foster the use of blind data extraction and analysis is a \textit{third-party service provider} that takes care of metric extraction and data analysis blindly. For example, the researchers conduct the experiment and, when needed, sanitize the raw dataset (\eg it removes any label to the treatments). Then they submit the raw dataset to this service provider, which extracts the metrics and then analyzes the data. After analyzing the data, the service provider sends the results to the researches. In this respect, R5 said:
\begin{mdframed}
\tablescript{An example could be an online service for data analysis where each participant, at the end of the [experimental] task, uploads its data on that platform and then someone else performs the data analysis. So who carries out the experiment does not interact with or manipulate the data, rather only acknowledges the results of the analysis. Clearly, this is costly and not easy to be~realized.
}\end{mdframed}
This strategy also has its drawbacks. As pointed out by R5, it is not easy to realize such a system. Also, the researchers should trust the service provider as well as the people that perform blindly the data extraction and analysis. Furthermore, it would most likely introduce extra costs. The third strategy consists of a \textit{guideline} for applying blind data extraction and analysis in SE. R6 told us:
\begin{mdframed}
\tablescript{Someone should try to give guidelines on how to put them [blind data extraction and analysis] in practice.
}\end{mdframed}
Finally, \textit{empirical evidence} on the effectiveness of blind data extraction and analysis in SE would foster the adoption of these blind techniques. On this matter, R4 said:
\begin{mdframed}
\tablescript{It would be nice if there could be some pilots or meta-studies that demonstrate how blind analysis and extraction change the results in either way, in favor or against the researcher's hypothesis. 
}\end{mdframed}

\textbf{Intention to Use Blind Data Extraction and Analysis.}
The interviewees stated they would take into account blind data extraction and analysis for their experiments. R8~stated:
\begin{mdframed}
\tablescript{If I have to participate in a large joint experiment between several research groups, I can take this into account when assigning the roles, why not! Instead of doing everything myself. 
}\end{mdframed}

\balance

\section{Discussion and Conclusion} \label{sec:discussion}
According to the interviewees, post-hoc hypotheses are not questionable as long as the researchers explicitly mention their use or it is possible to ground such hypotheses on prior work. Furthermore, this practice could be used to gain new insights into the investigated phenomenon. 
Similarly, the post-hoc outlier removal practice is not always questionable. In particular, it is considered acceptable when the researchers provide the results (after and before the outlier removal), justify the outlier removal, and discuss the causes behind possible differences.
This is in contrast with the guidelines for evaluating SE experiments by Kitchenham \etal~\cite{kitchenham2002preliminary} for which \textit{``the analysis protocol needs to address how drops out were handled''}. According to the authors, a clear outlier dropout analysis is particularly relevant for researchers interested in integrating the results of similar experiments (\eg meta-analysis).
In other words, from SE researchers' perspective, some QRPs are acceptable in certain cases, as recognized in previous studies (\eg~\cite{JLP12}). The question that arises is to what extent QRPs relates to the presence of RB in SE experiments. 

Based on our findings and those by Jorgensen~\etal \cite{Jorgensen:2016} and Shepperd \etal~\cite{Shepperd:14}, it seems that RB affects SE experiments.
Thus, we need to find solutions to mitigate RB as much as possible. Our results represent an initial exploratory step to establish guidelines to mitigate RB in SE~experiments based on solutions for SE researches and editorial/program boards.

\textbf{Solutions for Researchers}. Researchers can take into account blinding techniques to extract and analyze the data of their experiments. 
The importance of applying these techniques is central when performing meta-analyses~\cite{kitchenham2002preliminary}. The researchers we interviewed were favorable to use them (or at least to take them into account) in their future experiments. Although they acknowledged the usefulness of blind data extraction and analysis, such techniques alone do not solve the problem of RB but they are means to mitigate~it. 
Our findings suggested that blind data extraction and analysis are considered more effective in concealing information when the key roles (\eg study executor and data extractor) 
are covered by people that do not belong to the same research group.
Therefore, we encourage researchers to \textit{(i)}~involve external experts for blind data extraction and analysis or \textit{(ii)}~collaborate with other research groups to have external researchers taking care of data extraction and analysis.
However, it has emerged that involving external experts or collaborating with other research groups is not always possible.
Nevertheless, it is still possible to apply blind data extraction and analysis within the same research group.
For example, a simple form of blind data analysis can be achieved by relabelling the treatment groups in the dataset with non-identifying terms to hide the actual treatments from the data analyst. We recognize that in this case the analyst could guess the hidden information, but such a solution is surely better than having no blinding at all. To mitigate RB, the researcher could also consider using duplicate data analysis---\ie asking two or more people to analyze the data independently.
This technique could be easily extended to data extraction.
Duplicate data extraction and analysis could be applied as alternatives or in conjunction to blind data extraction and analysis.
Other two solutions to counteract RB, on the researcher side, are: \textit{(i)}~replicated experiments and \textit{(ii)}~means for improving the awareness of~RB.
Regarding the former, the underlying assumption of the interviewees is: the more the experimental results are confirmed, the lower the likelihood of RB is.
As for the latter solution, it is important to share knowledge on RB as well as strategies to deal with it---this paper represents a first step towards this~direction.

\textbf{Solutions for Editorial/Program Boards.}\label{sec:solBoards} The interviewees suggested fostering open data policies to mitigate RB since the more the studies are reproducible (\eg because datasets and analysis scripts are publicly available), the less the likelihood of biasing the results is.
In this respect, some conferences, such as ESEM 2018, have explicitly promoted open data policies. 
According to the interviewees, RB could be due to the behavior of some reviewers, namely  \textit{(i)}~their tendency to reject negative-results papers (\ie publication bias) or \textit{(ii)}~their tendency to focus too much on empirical assessment at the expense of contributions to the body of knowledge.
Accordingly, reviewers' behavior can lead researchers to bias their results to make their papers more publishable.
Therefore, acting on reviewers' behavior would possibly mitigate RB.  
The interviewees suggested ad-hoc tracks/issues for papers reporting negative results and studies having a weak empirical assessment but with a significant contribution to the body of knowledge.
For example, SANER 2018 has had a track where authors could submit negative-results papers, while the short paper track of EASE 2020 is fostering the submissions of research where a weak design could invalidate interesting findings. Tracks/issues for pre-registration papers is another solution to counteract the tendency to reject negative-results papers---\eg MSR 2020 is going to accept submissions of pre-registration papers.
Guidelines for reviewers can help mitigating publication bias as well.
For example, these guidelines should instruct reviewers not to judge papers based on study results (\eg positive/negative results).
Accordingly, editorial board should enforce reviews to comply with the guidelines.
Finally, conferences should consider having panels on RB in their programs to increase awareness of SE researchers about this problem and share solutions on how to limit it.



\section*{Acknowledgement}
This work was partially supported by: the KKS foundation through the S.E.R.T. Research Profile project at Blekinge Institute of Technology; the ``Digital Service Ecosystem'' project (Cod.
PON03PE-00136-1) funded by Italian MIUR; and ``Auriga2020'' project (Cod.
T5LXK18) funded by Apulia~Region.



\bibliographystyle{IEEEtran}
\bibliography{IEEEabrv,bibliography}

\begin{thebibliography}{10}
\providecommand{\url}[1]{#1}
\csname url@samestyle\endcsname
\providecommand{\newblock}{\relax}
\providecommand{\bibinfo}[2]{#2}
\providecommand{\BIBentrySTDinterwordspacing}{\spaceskip=0pt\relax}
\providecommand{\BIBentryALTinterwordstretchfactor}{4}
\providecommand{\BIBentryALTinterwordspacing}{\spaceskip=\fontdimen2\font plus
\BIBentryALTinterwordstretchfactor\fontdimen3\font minus
  \fontdimen4\font\relax}
\providecommand{\BIBforeignlanguage}[2]{{%
\expandafter\ifx\csname l@#1\endcsname\relax
\typeout{** WARNING: IEEEtran.bst: No hyphenation pattern has been}%
\typeout{** loaded for the language `#1'. Using the pattern for}%
\typeout{** the default language instead.}%
\else
\language=\csname l@#1\endcsname
\fi
#2}}
\providecommand{\BIBdecl}{\relax}
\BIBdecl

\bibitem{Ioannidis:2005}
J.~Ioannidis, ``Why most published research findings are false,'' \emph{PLOS
  Med}, 2005.

\bibitem{DAA08}
K.~Dwan, D.~G. Altman, and J.~A. Arnaiz, ``Systematic review of the empirical
  evidence of study publication bias and outcome reporting bias,'' \emph{PloS
  one}, 2008.

\bibitem{Wohlin:2012}
C.~Wohlin, P.~Runeson, M.~Hst, M.~C. Ohlsson, B.~Regnell, and A.~Wessln,
  \emph{Experimentation in Software Engineering}.\hskip 1em plus 0.5em minus
  0.4em\relax Springer, 2012.

\bibitem{Shepperd:14}
M.~Shepperd, D.~Bowes, and T.~Hall, ``Researcher bias: The use of machine
  learning in software defect prediction,'' \emph{IEEE Trans. Softw. Eng.},
  2014.

\bibitem{Jorgensen:2016}
M.~J{\o}rgensen, T.~Dyb\r{a}, K.~Liest{\o}l, and D.~I. Sj{\o}berg, ``Incorrect
  results in software engineering experiments: How to improve research
  practices,'' \emph{J. Syst. Softw.}, vol. 116, pp. 133--145, 2016.

\bibitem{MP15}
R.~MacCoun and S.~Perlmutter, ``Blind analysis: hide results to seek the
  truth,'' \emph{Nature}, vol. 526, no. 7572, pp. 187--189, 2015.

\bibitem{Karanicolas:2010}
P.~J. Karanicolas, F.~Farrokhyar, and M.~Bhandari, ``Blinding: Who, what, when,
  why, how?'' \emph{Can. J. Surg.}, vol.~53, no.~5, pp. 345--348, 2010.

\bibitem{Fucci:16}
D.~Fucci, G.~Scanniello, S.~Romano, M.~Shepperd, B.~Sigweni, F.~Uyaguari,
  B.~Turhan, N.~Juristo, and M.~Oivo, ``An external replication on the effects
  of test-driven development using a multi-site blind analysis approach,'' in
  \emph{Proc. of ESEM}.\hskip 1em plus 0.5em minus 0.4em\relax ACM, 2016, pp.
  3:1--3:10.

\bibitem{Sigweni:2015}
B.~Sigweni and M.~Shepperd, ``Using blind analysis for software engineering
  experiments,'' in \emph{Proc. of EASE}.\hskip 1em plus 0.5em minus
  0.4em\relax ACM, 2015, pp. 32:1--32:6.

\bibitem{King:2004}
N.~King, ``Using templates in the thematic analysis of text,'' in
  \emph{Essential Guide to Qualitative Methods in Organizational Research},
  C.~Cassell and G.~Symon, Eds.\hskip 1em plus 0.5em minus 0.4em\relax Sage,
  2004, pp. 256--270.

\bibitem{FCI17}
D.~Fanelli, R.~Costas, and J.~P.~A. Ioannidis, ``Meta-assessment of bias in
  science,'' \emph{Proc. of the National Academy of Sciences}, vol. 114,
  no.~14, pp. 3714--3719, 2017.

\bibitem{JLP12}
L.~John, G.~Loewenstein, and D.~Prelec, ``{Measuring the Prevalence of
  Questionable Research Practices With Incentives for Truth Telling},''
  \emph{Psychol. Sci.}, 2012.

\bibitem{Nuz15}
R.~Nuzzo, ``Fooling ourselves,'' \emph{Nature}, vol. 526, no. 7572, pp.
  182--185, 2015.

\bibitem{PWD10}
C.~J. Pannucci and E.~G. Wilkins, ``Identifying and avoiding bias in
  research,'' \emph{Plastic and reconstructive surgery}, vol. 126, no.~2, pp.
  619--625, 2010.

\bibitem{Nosek:2014}
B.~Nosek and D.~Lakens, ``Registered reports: a method to increase the
  credibility of published results,'' \emph{Social Psychology}, vol.~45, no.~3,
  pp. 137--141, 2014.

\bibitem{Miller:2011}
L.~Miller and M.~Stewart, ``The blind leading the blind: Use and misuse of
  blinding in randomized controlled trials,'' \emph{Contemporary Clinical
  Trials}, vol.~32, no.~2, 2011.

\bibitem{Page:2013}
S.~J. Page and A.~C. Persch, ``Recruitment, retention, and blinding in clinical
  trials,'' \emph{Am. J. Occup. Ther.}, vol.~67, no.~2, pp. 154--161, 2013.

\bibitem{Guest:2006}
G.~Guest, A.~Bunce, and L.~Johnson, ``How many interviews are enough?: An
  experiment with data saturation and variability,'' \emph{Field Methods},
  vol.~18, no.~1, pp. 59--82, 2006.

\bibitem{Thurmond:2001}
V.~Thurmond, ``The point of triangulation,'' \emph{J. Nurs. Scholarsh.}, 2001.

\bibitem{kitchenham2002preliminary}
B.~A. Kitchenham, S.~L. Pfleeger, L.~M. Pickard, P.~W. Jones, D.~C. Hoaglin,
  K.~El~Emam, and J.~Rosenberg, ``Preliminary guidelines for empirical research
  in software engineering,'' \emph{IEEE Trans. Softw. Eng.}, vol.~28, no.~8,
  2002.

\end{thebibliography}

\end{document}